\documentclass[usenatbib]{mn2e}
\usepackage{times}
\usepackage{epsf}
\usepackage{url}

\newcommand*{\D}{\mathop{}\!\mathrm{d}}
\newcommand*{\sign}{\mathop{\mathrm{sign}}\nolimits}
\newcommand*{\hnull}{\ensuremath{H_{0}}}
\newcommand*{\lnull}{\ensuremath{\lambda_{0}}}
\newcommand*{\onull}{\ensuremath{\Omega_{0}}}
\newcommand*{\knull}{\ensuremath{K_{0}}}
\newcommand*{\apriori}{a priori}
\newcommand*{\eg}{e.g.}
\newcommand*{\etc}{etc}
\newcommand*{\ie}{i.e.}
\newcommand*{\perse}{per se}
\newcommand*{\ack}{Acknowledgments}
\newcommand*{\Eqn}{Equation}

\newcommand*{\Fig}{Fig.}
\newcommand*{\Sect}{Section}
\newcommand*{\Sects}{Sections}
\newcommand*{\eqn}{equation}

\newcommand*{\fig}{fig.}
\newcommand*{\tab}{table}
\newcommand*{\sect}{section}
\newcommand*{\ndash}{--} 
\newcommand*{\pdash}{ -- } 

\title[The flatness problem in classical cosmology]%
{Is there a flatness problem in classical cosmology?}

\author[P.~Helbig]%
{Phillip Helbig\thanks{E-mail: helbig@astro.multivax.de}\\
Thomas-Mann-Str.~9, D-63477 Maintal, Germany}

\begin{document}

\date{Accepted 2011 December 05.  Received 2011 December 04;
in original form 2011 November 27}

\pagerange{\pageref{firstpage}--\pageref{lastpage}} \pubyear{2012}

\maketitle
\label{firstpage}

\begin{abstract}
I briefly review the flatness problem within the context of classical
cosmology and examine some of the debate in the literature with regard
to its definition and even the question whether it exists.  I then
present some new calculations for cosmological models which will
collapse in the future; together with previous work by others for models
which will expand forever, this allows one to examine the flatness
problem quantitatively for all cosmological models.  This leads to the
conclusion that the flatness problem does not exist, not only for the
cosmological models corresponding to the currently popular values of
\lnull\ and \onull\ but indeed for all Friedmann\ndash Lema\^{\i}tre
models. 
\end{abstract}


\section{Introduction}

The flatness problem has been called one of the outstanding puzzles in
cosmology \citep[\eg][]{RDickePPeebles79a}.  This in itself is rather
puzzling in view of the fact that the arguments in favour of it being a
problem are rather vague and heuristic, while quantitative arguments
have been presented against the claim that it is a problem, at least for
some classes of cosmological models
\citep[\eg][]{PColesGEllis97a,KLake05a}.  The flatness problem is one of
the main motivations for the inflationary scenario \citep{AGuth81a}. Of
course, if there is no flatness problem (or, indeed, even if there were
no motivation at all for inflation), this does not mean that inflation
could not have occurred.  However, it does mean that inflation should
not be taken as given based on the belief that it explains away the
flatness problem and thus without it classical cosmology leads to absurd
conclusions. 

The plan of this paper is as follows.  In \Sect~\ref{cosmo} I present
the basic equations needed in the rest of the paper, mainly to define my
notation (unfortunately, there is not a uniform notation in the
literature) and give an overview of the entire cosmological parameter
space relevant to the discussion.  Section~\ref{history} gives a brief
historical overview of the flatness problem and some qualitative
arguments against it.  In \Sect~\ref{collapse} I discuss a new
quantitative argument regarding cosmological models which will collapse
in the future.  Sections~\ref{Lake} and~\ref{anthro} discuss previous
quantitative results by others for other classes of cosmological models.
Section~\ref{summary} summarizes the results for all cosmological
models.

\section{Basic cosmology}
\label{cosmo}

I assume that, at the level of detail necessary, the universe can be
described by the Friedmann\ndash Lema\^{\i}tre equation 
\begin{equation}
\label{fl}
\dot R^2 = \frac{8\pi G\rho R^2}{3} + \frac{\Lambda R^2}{3} - kc^{2}
\end{equation}
with the dimensionless constant $k$ equal to $-1$, $0$, $+1$ depending
on spatial curvature (negative, vanishing or positive, respectively); $R$
is the scale factor (with dimension length) of the universe, $G$ the
gravitational constant, $\rho$ the density, $\Lambda$ the cosmological
constant (dimension time$^{-2}$) and $c$ the speed of light.  It is
useful to define the following quantities: 
\begin{displaymath}
\begin{array}{rlllll}
H       & :=  &  \frac{\dot R}{R}  & & & \\
& & & & & \\
\lambda & :=  &  \frac{\Lambda}{3H^{2}} & & &\\
& & & & & \\
\Omega  & :=  &  \frac{\rho}{\rho_\mathrm{crit}} 
              &  \equiv & \frac{8\pi G\rho}{3H^{2}}
& \\& & & & & \\
K & := & \Omega + \lambda - 1 & & & \\
& & & & & \\
q       & :=  &  \frac{-\ddot R R}{\dot R^{2}} & \equiv &
                                                  \frac{-\ddot R}{RH^{2}} &
\equiv \frac{\Omega}{2} - \lambda
\end{array}
\end{displaymath}
which are all dimensionless except that $H$ has the dimension
time$^{-1}$.  $H$ is the Hubble constant, $\lambda$ the normalized
cosmological constant, $\Omega$ the density parameter, $k=\sign(K)$ and
$q$ is the deceleration parameter.\footnote{$q>0$ implies that the
universe is \emph{de}celerating; the minus sign is included in the
definition since at the time of its invention it was assumed that
$\lambda=0$ which implies that $q>0$ (or $q=0$ in the case of a universe
devoid of matter).}  For $\lambda=0$ and $k=0$, $\rho =
\rho_{\mathrm{crit}} = \frac{3H^{2}}{8\pi G}$.  This density is
`critical' in the sense that, for $\lambda=0$, a greater (lesser)
density implies a positive (negative) curvature and a universe (assumed
to be expanding now) which will collapse in the future (expand forever);
similarly, for $k=0$, a greater (lesser) density implies a negative
(positive) cosmological constant and a universe (assumed to be expanding
now) which will collapse in the future (expand forever).  However, in
the general case ($k \neq 0$ and $\lambda \neq 0$), $\rho_\mathrm{crit}$
doesn't have any special meaning, though $\Omega$ remains a useful
parameter.  \Eqn~(\ref{fl}) can be rearranged, using the definitions
above, to give 
\begin{equation}
\label{R}
R = \frac{c}{H}\frac{\sign(K)}{\sqrt{|K|}} ,
\end{equation}
thus $R$ is positive for $k=+1$ and negative for $k=-1$; for $k=0$, $R$
can be defined as $\frac{c}{H}$. 

It can be useful to express \eqn~(\ref{fl}) with the values of the
dimensionless parameters as observed now, denoted by the suffix $0$.
This leads to 
\begin{equation}
\label{fl-0}
\dot R^2 = \dot R_0^2 \left( \frac{\onull R_0}{R} +
           \frac{\lnull R^2}{R_0^2} - K_0 \right)
\end{equation}
or, making use of the definition of $H$,
\begin{equation}
\label{fl-H}
H^2 = \hnull^2\left( \frac{\onull R_0^3}{R^3} +
           \lnull - \frac{K_0 R_0^2}{R^2} \right)  .
\end{equation}
(Note that, with 
\begin{equation}
z := \frac{R_0}{R} - 1
\end{equation}
this leads to 
\begin{equation}
\frac{\D z}{\D t} = \frac{\D z}{\D R}\frac{\D R}{\D t} =
\frac{\D z}{\D R}\dot R =
- \hnull(1 + z)\sqrt{F(z)}
\end{equation}
where
\begin{equation}
F(z) = \onull(1+z)^3 - K_0(1+z)^2 +\lnull
\end{equation}
which is the starting point for calculating light-travel time and
distance as a function of redshift.)  In general, $H$, $\lambda$ and
$\Omega$ all change with time.  Note that since 
\begin{equation}
\label{lambda}
\lambda = \lnull \left(\frac{\hnull}{H}\right)^2  ,
\end{equation}
the change in $\lambda$ with time is due entirely to the change in $H$
with time, since $\Lambda$ is constant.  Also, since the density $\rho$
is inversely proportional to the cube of $R$, 
\begin{equation}
\label{Omega}
\Omega = \onull \left(\frac{\hnull}{H}\right)^2
\left(\frac{R_0}{R}\right)^3                            ,
\end{equation}
the variation in $\Omega$ is due both to variation in $H$ and to the
decrease in density as the universe expands. 

\begin{figure}
\epsfxsize=\columnwidth
\epsfbox[32 32 496 496]{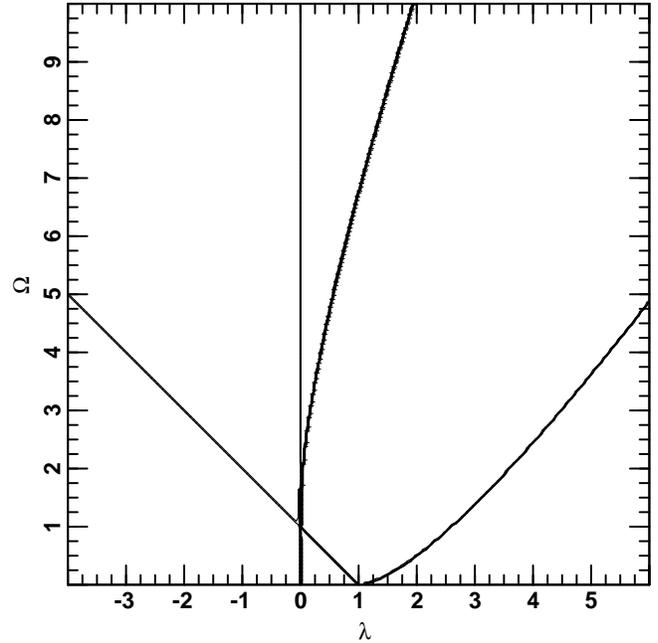}
\caption{The $\lambda$-$\Omega$ plane.  See text for details.}
\label{overview}
\end{figure}

\begin{figure}
\epsfxsize=\columnwidth
\epsfbox[32 32 496 496]{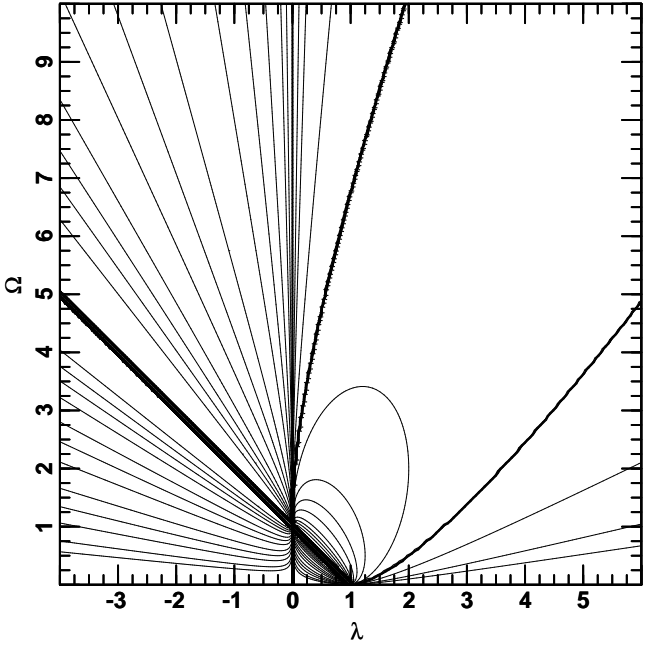}
\caption{Evolutionary trajectories in the $\lambda$-$\Omega$ plane.}
\label{trajectories}
\end{figure}

In general, $\lambda$ and $\Omega$ evolve with time.  (They do not for
the ($\lambda$,$\Omega$) values of $(0,0)$ (Milne model), $(0,1)$
(Einstein\ndash de~Sitter model), $(1,0)$ (de~Sitter model) and for the
static Einstein model [in which $\lambda$  and $\Omega$ are infinite
(though $\Lambda$ and $\rho$ are not; $\Lambda = 4\pi G\rho$) since $H$
is 0].)  For an excellent discussion of the evolution of $\lambda$ and
$\Omega$ (though expressed in the older notation using
$\sigma=\frac{\Omega}{2}$ and $q=\sigma-\lambda$), see
\citet{RStabellSRefsdal66a}.  For a classification based on the
evolution of $\lambda$ and $\Omega$ using the more modern notation used
in the present paper, see \tab~1, \fig~1 and \sect~a.i.\ in
\citet{Phelbig96a}.\footnote{This is the arXiv version of
\citet*{RKayserHS97a} but in contrast to the A\&A paper contains the
User's Guide as well; also available at
\url{http://www.astro.multivax.de:8000/helbig/research/p/ps/angsiz_guide.ps}.} For the present discussion,
the basic information needed can be seen in \Fig~\ref{overview} [for
derivation, see \citet{RStabellSRefsdal66a}] which shows an overview of
the $\lambda$-$\Omega$ plane. 
The vertical line corresponds to $\lambda=0$; the diagonal line
corresponds to $k=0$ with $k=-1$ below it and $k=+1$ above it. The curve
near the vertical line [corresponding to the A1 curve in
\citet{RStabellSRefsdal66a}] separates models which will collapse (to
the left) from those which will expand forever (to the right).  Models
on the curve start arbitrarily close to the Einstein\ndash de~Sitter
model (like all non-empty big-bang models) and asymptotically approach
the static Einstein model which has $\lambda=\Omega=\infty$ (since
$H=0$; $\Lambda$ and $\rho$ have finite values).  The other curve
[corresponding to the A2 curve in \citet{RStabellSRefsdal66a}] separates
big-bang models (to the left) from non--big-bang models (to the right);
the latter contract from an infinite to a finite size then expand
forever.  Models on the curve start at the static Einstein model and
asymptotically approach the de~Sitter model (the latter feature is
shared with all models which expand forever and have $\lambda>0$). 
\Fig~\ref{trajectories} shows some sample trajectories in the
$\lambda$-$\Omega$ parameter space superposed in \Fig~\ref{overview}. 
Note that all the lines and curves in \Fig~\ref{overview} correspond to
trajectories. In addition, there are models at the intersections of the
lines with ($\lambda,\Omega$ ) values of $(0,0)$ (the Milne model),
$(0,1)$ (the Einstein\ndash de~Sitter model), $(1,0)$ (the de~Sitter
model) and at $(\infty,\infty)$ (the Einstein model); in these, $\lambda$
and $\Omega$ are constant in time. Also, note that the trajectories do
not cross; this means that the history of a cosmological model (\ie~the
way $\lambda$ and $\Omega$ change with time) is completely determined by
the values at any point on it (in practice, by measuring the values at
the present time, \lnull\ and \onull).  Since the lines and curves are
also valid trajectories, this means that the signs of $\lambda$ and $K$
cannot change and that a model with $\Omega=0$ at any time has
$\Omega=0$ at all times. 

\begin{figure}
\epsfxsize=\columnwidth
\epsfbox[32 32 496 496]{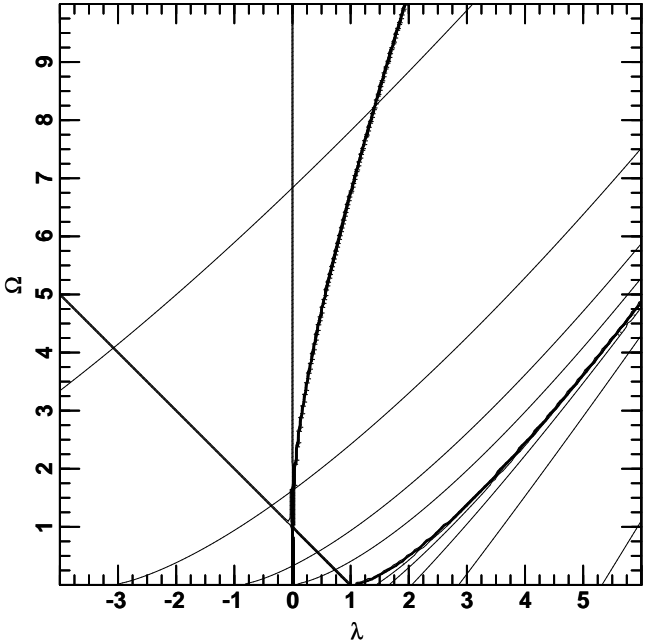}
\caption{The age of the universe in units of $H^{-1}$.  From upper left 
to lower right, contours are at 0.4, 0.6, 0.8, 1.0, 1.0, 0.8, 0.6, 0.4 
and 0.2  Between the two contours at 1.0 is the A2 curve which 
corresponds to $\infty$.  To the left of the curve the contours indicate 
the time since the big bang; to the right they indicate the time since 
the universe started expanding from its minimum size.}
\label{age}
\end{figure}

\Fig~\ref{age} shows contours of constant $Ht$, \ie~the age of the universe
in units of the Hubble time ($H^{-1}$). 
The relation between these contours and the trajectories shown in 
\Fig~\ref{trajectories} is
important for the discussion in \Sects~\ref{collapse} and \ref{anthro}. 
Note that this is a smooth and well behaved function of $\lambda$ and
$\Omega$, independent of the geometry (global curvature), origin and
fate (big bang or not, recollapse or eternal expansion) or contents
(matter, cosmological constant) of the universe (except of course that
the age of the universe is infinite on the A2 curve; contours to the
right of the A2 curve indicate the time since the minimum of the scale
factor).

\section{A brief history of the flatness problem}
\label{history}

The flatness problem appears in two forms.  One states that if
$\Omega\approx 1$ today, then in the early universe it was arbitrarily
close to 1; the assumption is that some `mechanism' is needed to explain
this `fine-tuning' \citep[\eg][]{AGuth81a}.  (It is usually not stated
but almost always assumed that no fine-tuning would be necessary if
$\Omega$ were not $\approx 1$ today.)  The other states that if $\Omega$
changes with time, then we should be surprised that $\Omega$ is (still)
$\approx 1$ today \citep[\eg][]{KLake05a}.  In other words, the problem
is that the mechanism whereby $\Omega$ is forced to be $\approx 1$ is
unknown.  Solving one of these variants of the flatness problem does not
necessarily solve the other variant. 

[Historically, the flatness problem was first discussed during a time
when $\lambda$ was thought to be zero.  Thus, most discussions took this
as given.  As mentioned above, in this case $\Omega=1$ corresponds to
$k=0$, \ie~a flat universe, hence the name `flatness problem' for the
question why the universe is (nearly) flat today considering that
$\Omega$ evolves away from $\Omega=1$ ($k=0$) with time.  If $\lambda$
is not constrained to be zero, then the flatness problem should be
re-phrased as the Einstein\ndash de~Sitter problem, \ie~the question is
why the universe is (in some sense) close to the Einstein\ndash
de~Sitter model (which is an unstable fixed point and a repulsor) today
when $|\lambda|$ and $\Omega$ can take on values between $0$ and
$\infty$. However, for brevity I will continue to use the term 
`flatness problem'
even for the more general case and sometimes mention only the change in
$\Omega$ with time.  Nevertheless, it is important to keep
in mind (though it doesn't change the thrust of the argument in all
cases) that the discussion should really be about the Einstein\ndash
de~Sitter problem.]

One possible `solution' to the flatness problem is simply to claim that
$\Omega\equiv 1$.  In this case (for $\lambda=0$), since $\Omega$
doesn't evolve with time, no explanation is needed as to why
$\Omega\approx 1$ \emph{today}; in other words, we don't have to worry
about living at a special time.  Of course, the fact that $\Omega$ just
\emph{happens} to be 1 is thought by some to be improbable in some
ill-defined sense, so it is more convincing if there is some mechanism,
such as inflation, which produces $\Omega=1$ to at least a very good
approximation, rather than having to rely on the lucky coincidence of
living in a cosmological model in which $\Omega$ does not change with
time.  (Of course, the Einstein\ndash de~Sitter model is now ruled out
by observations.  Before this was the case, the argument `$\Omega$ must
be exactly $1$ because if not it will evolve away to to an arbitrarily
large or arbitrarily small value' was indeed used.)  However, this
solution does not work in practice, as has been pointed out by
\citet{AGuth81a} and \citet{PColesGEllis97a}: even if our universe is
described by the Einstein\ndash de~Sitter model when averaged over large
scales, this is obviously not the case on smaller scales; any deviations
from the Einstein\ndash de~Sitter model would then grow with time, even
if the entire universe were still, on average, described by the
Einstein\ndash de~Sitter model, so that we would not expect our
observable universe to be described exactly by the Einstein\ndash
de~Sitter model at a `random' time. 

Note that a flat universe ($k=0$, $\Omega+\lambda=1$) doesn't really
offer much of an advantage.  Though it is true that if
$\Omega+\lambda=1$ then this equation always holds, while in general the
sum is time-dependent, the same argument can be applied to the value of
$\Omega$ in the flat case as in the case of $\lambda=0$.  In other
words, although it is true that in a flat universe $\Omega + \lambda$
doesn't change with time, $\Omega$ itself suffers from the same
`problem' it does in the $\lambda=0$ case (except in the special cases
of the Einstein\ndash de~Sitter model and the de~Sitter model). Thus,
for the flatness problem in the broader sense of the term
(\ie~understanding why $\Omega\approx 1$ today), the cosmological model
favoured by observations ($\lnull\approx 0.7$, $\onull\approx 0.3$,
$k=0$ to very good precision) is actually worse than the Einstein\ndash
de~Sitter model, even if it were exactly flat. 

Another solution discussed by \citet{PColesGEllis97a} is to deny that it
is a problem that we live at some special time in the history of the
universe, since it is not equally likely that we could live at any time;
this is of course the weak anthropic principle.  While the weak
anthropic principle certainly plays some role in determining the
probability of observing given values of the cosmological parameters, it
is not obvious that it can quantitatively explain why our universe is so
close to the Einstein\ndash de~Sitter model.  In any case, most people
would probably prefer an explanation which doesn't rely on the anthropic
principle.  I shall return to the weak anthropic principle in
\Sect~\ref{anthro}.  [Note that a perfectly spatially flat universe is
always perfectly spatially flat, and thus does not suffer from the
`special time' problem with regard to flatness (though one might need
some explanation for the perfect flatness itself).  However, it still
suffers from the Einstein\ndash de~Sitter problem (except in the cases of
the Einstein\ndash de~Sitter universe and the de~Sitter universe in
which there is no evolution of $\lambda$ and $\Omega$).]

One definition of a `special time' is a time such that the radius of
curvature is comparable to the distance to the (particle or event)
horizon or the radius of the Hubble sphere.  Since in general these
quantities are not simply proportional and evolve differently with time,
it could be seen as a coincidence (possibly needing explanation) if they
are comparable.  A perfectly flat universe with an infinite radius of
curvature doesn't have this problem since the radius of curvature is
always infinite.  However, this `advantage' of a flat universe is not so
much a solution to the time-scale problem as a statement that it doesn't
exist (with respect to the evolution of the curvature radius) in such a
universe; one could still ask what `causes' the universe to be spatially
flat. Also, as noted above, a spatially flat universe still suffers from
the `Einstein\ndash de~Sitter problem'.  (This coincidence problem is
similar to, but distinct from, the flatness problem.  If at all, it
exists only for $\lambda=0$ and $\Omega<1$.  I plan to discuss this in a
future paper.  As is shown below, this class of models is somewhat
atypical with respect to the flatness problem; this is also the case for
the coincidence problem.  Since $\lambda=0$ was a common assumption when
these problems were first discussed, sometimes particular features of
this class of models are mistaken for generic features of
Friedmann-Lema\^{\i}tre models.) 

These three `solutions' to the flatness problem\pdash $\Omega \equiv 1$
(and $\lambda\equiv 0$), $k=0$, anthropically selected special time\pdash
are thus unsatisfactory. Are there any satisfactory ones?

\subsection{The qualitative flatness problem: Is a fine-tuning of 
initial conditions required?}
\label{qualitative}

The flatness problem is often presented as a fine-tuning problem
\citep[\eg][]{AGuth81a}: if $\Omega$ is near 1 to day, then at some time
$t_{\mathrm{fine}}$ in the past it must have been 1 to a very high
accuracy.  I refer to this sense of the flatness problem as the
`qualitative flatness problem'.  This argument is completely bogus, as
has been pointed out by many authors
\citep[\eg][]{PColesGEllis97a,KLake05a}: \emph{all} non-empty models
begin their evolution at the Einstein\ndash de~Sitter model, so of
course the further back in time one goes, the `more finely tuned'
$\Omega$ is.  The point is, within the context of classical cosmology,
there is nothing special about a time $t_{\mathrm{fine}}$ chosen so that
$\Omega$ is very close to 1 at that point.  Times such as the Planck
time are often used in examples of the flatness problem, but not only is
the Planck time irrelevant within purely classical cosmology but also
there is no known theory which predicts the likelihood distribution of
$\Omega$ values at this (or any other) time.  This should be obvious
from the fact that for \emph{any} value of $\Omega$ today, one can
choose a time $t_{\mathrm{fine}}$ such that $\Omega$ is as close to 1 as
desired.  In other words, the flatness `problem' would still exist if
$\Omega$ were appreciably greater or less than 1 today, only the time
$t_{\mathrm{fine}}$ for a given degree of fine tuning would occur at an
earlier time.  (Why this is less mysterious for many people is not clear 
to me.)  Thus, the `problem' still remains.  (Also, the degree of
fine tuning required for there to be a `problem' is not well defined,
but is arbitrary and subjective.)  Alternatively, if the problem is seen
as a problem connected with the observed value of $\Omega$, then the
fact that it exists for many other values as well, indeed for all
values, could lead to the conclusion that it is not a problem at all.
This is thus a qualitative argument against the existence of the
qualitative flatness problem. 

\citet{GEvrardPColes95a} \citep[see also][]{PColesGEllis97a} also point
out that the assumption implicit in the qualitative flatness problem,
namely that some wide range of $\Omega$ values are \apriori\ equally likely at
some early time, constitutes a prior which is incompatible with the
assumption of minimal information.  This can be regarded as a
quantitative solution to the qualitative flatness problem (or, perhaps,
an argument against its existence). 

The qualitative flatness problem thus does not exist; it is merely a
consequence of the way in which a universe, described by the
Friedmann\ndash Lema\^{\i}tre equation, evolves and how dimensionless
observable quantities such as $\Omega$ are defined.  Suppose one
is standing at the bottom of the famous Leaning Tower of Pisa and
observes a cannonball dropping past one's face to the ground.  In some
absolute sense it is travelling slowly (its speed is much less than the
speed of light, say).  One can measure the acceleration and calculate
that, at a time $t_{\mathrm{fine}}$ in the past, its velocity must have
been extremely `finely tuned' to almost zero.  In fact, at a finite time
in the past, its velocity \emph{was} zero, and at that time it was at a
particular place (say, the top of the tower).  What's more, other
objects which, due to the effects of air resistance, are falling past
one at other speeds are \emph{all} found to be `finely tuned' so that
their velocity was 0 at a given time in the past (and for all objects,
the height at that time is the same). 

Obviously, there is nothing at all puzzling about this scenario.
Nevertheless, this is the type of `evidence' which is often presented
for the `existence of the flatness problem', with $\Omega$ taking the
place of velocity (being 1 at the beginning and not 0, of course).
Argument from analogy can be misleading, so one should put more faith in
quantitative arguments. 

In the above case, the solution is obviously to be found in the initial
conditions: Galileo is dropping objects from the top storey.  In fact,
in the cosmological case the lack of a need for fine-tuning is even more
obvious.  While Galileo could give the objects an initial velocity
rather than just dropping them, any non-empty big-bang universe
described by the Friedmann\ndash Lema\^{\i}tre equation always begins
arbitrarily close to the Einstein\ndash de~Sitter model.  There is no
need for fine-tuning since there is no possible range of values at the
initial time.  
See \Sect~\ref{quantitative} for the discussion of a somewhat better,
but still flawed, analogy.  Nevertheless, even if it is not a puzzle
why $\Omega=1$ at early times, one can still ask whether we should be
surprised that $\Omega\approx 1$ today.

\subsection{The quantitative flatness problem: Should we be surprised
that $\onull\ \approx 1$?}
\label{quantitative}

What is the relation between the two forms of the flatness problem
discussed above?  Within the context of classical cosmology, the first
statement, that the universe was arbitrarily close to the Einstein\ndash
de~Sitter model near the beginning, is almost always true.  (It
is always true for non-empty big-bang models.  For empty models, similar
arguments apply: For $\lambda<1$, the starting point is not the
Einstein\ndash de~Sitter model but rather the Milne model
($\lambda=\Omega=0$).  For $\lambda>1$ the starting point is the
de~Sitter model ($\lambda=1$ and $\Omega=0$), although these are not
big-bang but rather bounce models.  (The limiting case of the de~Sitter
model itself can be thought of as a big-bang model in which the big
bang occurred in the infinite past.)  Another non--big-bang model is the
static Einstein model.  At first sight, this model seems completely
different than the Einstein\ndash de~Sitter model.  However,
mathematically both are unstable fixed points.  Interestingly,
\citet{AEddington30a} argued that this is a mark against the
Einstein model, since it is unstable (thus unlikely to hold in a
realistic universe): the attraction due to gravity and the repulsion due
to the cosmological constant are exactly balanced.  Since the density of
matter decreases if the universe expands and vice versa, it is unstable
not only to changes in the value of the cosmological constant or the
density \perse, but also to departures from the static state.  The
Einstein\ndash de~Sitter universe is unstable in exactly the same sense:
as long as it is not perturbed, the values of the cosmological
parameters remain constant; if they are even slightly different, they
evolve away from the pure Einstein\ndash de~Sitter state.  Of course, in
some sense this is not a problem since if we assume that the universe is
exactly described by the Friedmann\ndash Lema\^{\i}tre equation, then
the behaviour at all times is fixed; there is no way to perturb it.  On
the other hand, if only because the universe is not completely
homogeneous, the Friedmann\ndash Lema\^{\i}tre equation cannot be exact,
so the objection is a valid one.  Interestingly, the same argument is
rarely used against the Einstein\ndash de~Sitter universe, even though
mathematically both are unstable fixed points and hence suffer from the
same weakness.  [In fact, the argument is turned around: \emph{because}
the Einstein\ndash de~Sitter universe is unstable, if the observational
constraints on the cosmological parameters describing our universe are
compatible with the Einstein\ndash de~Sitter model, it is much more
probable that our universe is described (almost) exactly by the
Einstein\ndash de~Sitter model, rather than one nearby in the parameter
space.  Of course, the caveats regarding `probable', `almost' and
`nearby' mentioned above make this an invalid line of argument.]  In
both cases, a universe described exactly by the model in question would
be stable (but would contain no cosmologists, since these presumably
require inhomogeneities, and of course could not be perturbed from
`outside the universe') while a realistic universe would be at least
locally unstable even if the model were an accurate description of a
large-scale average.  Of course, the static Einstein universe was ruled
out observationally before the Einstein\ndash de~Sitter universe was,
but while observational arguments themselves are essential, they should
not influence what should be purely theoretical or mathematical
arguments.)  As discussed above, this can be interpreted to mean that
this aspect of the flatness problem exists for all cosmological models
(\ie~regardless of what values we observe for \lnull\ and \onull), which
means that in this respect there is nothing special about our universe,
or that it is not a problem at all, but just a consequence of
definitions.  Obviously, the second form of the problem (If $\Omega$
evolves, even to arbitrarily large values in a finite time, should we be
surprised if it is $\approx 1$?) holds only if $\Omega\approx 1$ today;
on the other hand, holding that the first form is not a problem (which,
as outlined above, I am not alone in claiming) does not automatically
solve the second problem.  The rest of this paper is concerned mainly
with the second form: should we be surprised that $\Omega\approx 1$
today?  This `quantitative flatness problem' is more subtle, but also
has solutions within the context of classical cosmology. 

An analogy often used in discussing the flatness problem is that of a
tightrope walker \citep[\eg][]{PColesGEllis97a,PColes09a}:  The
Einstein\ndash de~Sitter model corresponds to the balanced tightrope
walker.  As long as he stays balanced, he will stay where he is.
However, a slight deviation will grow, quickly bringing the tightrope
walker far from the tightrope.  If one walks by a circus tent at an
arbitrary time and looks inside, one would expect to find the tightrope
walker either balanced on the tightrope or on the ground after having
fallen; it is extremely improbable that one would just happen to see him
during his fall.  The ground corresponds to one of the extreme values of
$\Omega$ ($\infty$ or $0$).  Every statement of the quantitative
flatness problem essentially follows this example.  Were it a valid
analogy, then there would indeed be a quantitative flatness problem.  In
the following sections, I will show quantitatively why the analogy is
wrong and hence why there is no quantitative flatness problem.

\section{Cosmological models which collapse in the future: a new 
solution to the flatness problem}
\label{collapse}

All cosmological models (assumed to be expanding now) with $\lambda<0$
will collapse in the future: $\ddot R$ is negative for all values of $R$
and for large $R$ is proportional to $R$.  Models with $\lambda=0$ will
collapse for $\Omega>1$.  In addition, models with $\lambda>0$ will
collapse provided that $\Omega>1$ (which in this case implies $K>0$,
\ie~$k=+1$), $q>0$ and $\alpha<1$, where 
\begin{equation}
\label{alpha}
\alpha = \sign(K)\frac{27\Omega^{2}\lambda}{4K^{3}}
\end{equation}
\citep{RStabellSRefsdal66a,KLake05a}.  (The A1 and A2 curves mentioned
above have $\alpha=1$.)  In \Fig~\ref{overview}, these are in the area
between $\lambda=0$ and the A1 curve.  Empty big-bang models start
arbitrarily close to the Milne model with ($\lambda$,$\Omega$) values of
$(0,0)$; non-empty big-bang models start arbitrarily close to the
Einstein\ndash de~Sitter model with ($\lambda$,$\Omega$) values of
$(0,1)$.  The evolution of $\lambda$ and $\Omega$ can be viewed as
trajectories in the parameter space: $\lambda$ and $\Omega$ evolve from
the starting point to infinity and return along the same trajectory.
[For the definitive discussion, see \citet{RStabellSRefsdal66a}; a very
useful visualisation can be found at \citet{JLeahy03a}.]  Note that such
trajectories do not intersect; this means that a trajectory is uniquely
determined by measuring $\lambda$ and $\Omega$ at any one time
(practical is of course doing so now). 

The fact that $\lambda$ and $\Omega$ evolve to $\infty$ (and back) in a
finite time immediately illustrates what is wrong with the
tightrope-walker analogy for these models: the proper analogy for a
universe which will collapse in the future would be a tightrope walker
who, if he falls off the tightrope, doesn't stop when he hits the
ground, but rather continues through the Earth and to infinity and back,
finally approaching the tightrope from below until he is ultimately 
almost balanced again.  It should be immediately obvious that in this
case all values of $\lambda$ and $\Omega$ cannot be equally probable. 
Making the analogy more quantitative, it turns out that the tightrope
walker actually spends most of his time between the rope and the floor,
thus we should \emph{not} be surprised to find him somewhere between
the tightrope and the ground when we look inside the tent.  In other
words, there is no quantitative flatness problem in these models. 

\begin{figure}
\epsfxsize=\columnwidth
\epsfbox[32 32 496 496]{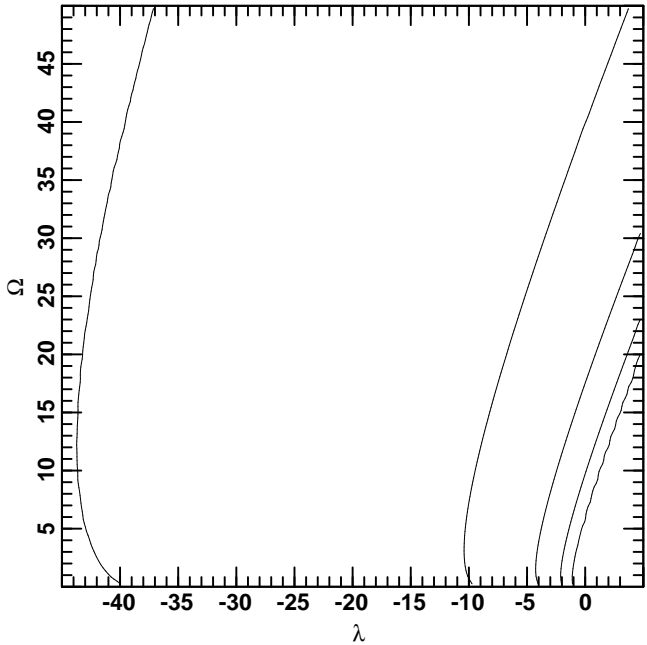}
\caption{The age of the universe as a fraction of the time between the 
big bang and maximum expansion.  Contours, from right to left, are at
0.5, 0.6, 0.7, 0.8 and 0.9.} 
\label{quotient1}
\end{figure}

To quantify this, I have calculated the quotient of the age of the
universe now and at the time of maximum expansion as a function of
$\lambda$ and $\Omega$.  The age of the universe is given by 
\begin{equation}
\label{age-eqn}
t = \int\limits_{0}^{R(t)}\frac{\D R}{\sqrt{\dot R_0^2 
     \left( \frac{\onull R_0}{R} + \frac{\lnull R^2}{R_0^2} - 
     \knull \right)}}
\end{equation}
which follows from \eqn~(\ref{fl-0}).
For the
current age, the upper limit is given by \eqn~(\ref{R}); at the time of
maximum expansion it is found by calculating the (smallest) zero of
$\dot R^{2}$ (since $\dot R^{2}$ cannot be negative).  This is shown in
\Fig~(\ref{quotient1}). 
It is clear that large values of $\lambda$ and $\Omega$ occur only
during a relatively short time in the history of the universe, near the
time of maximum expansion (at the precise time of maximum expansion,
$\lambda$ and $\Omega$ are infinite since $H=0$). 
\Fig~5 shows the same for $\lambda>0$.  
\begin{figure}
\epsfxsize=\columnwidth
\epsfbox[32 32 496 496]{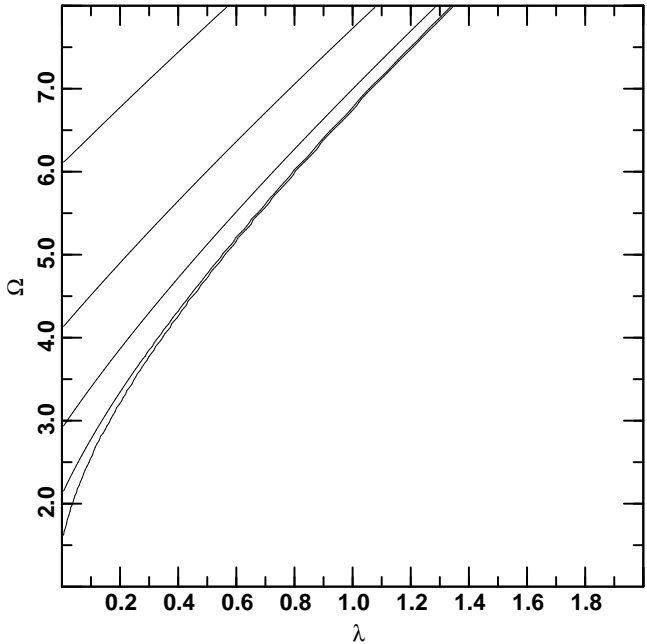}
\caption{The age of the universe as a fraction of the time between the 
big bang and maximum expansion.  Contours, from the upper left, are at
0.5, 0.4, 0.3, 0.2 and 0.1.} 
\label{quotient2}
\end{figure}
While the constraints aren't quite as strong here, in the
\Sect~\ref{Lake} I show that a different argument renders many of the
models in this part of parameter space unlikely. 

Note that this argument is completely independent of \hnull: whatever
the value of \hnull, \ie~whatever the age of the universe, models which
recollapse have large values of $\lambda$ and $\Omega$ for only a
relatively short time.

\section{Cosmological models which expand forever: reversing the 
fine-tuning problem}
\label{Lake}

An interesting point has been made by \citet{KLake05a}, though it is
implicit in \citet{RStabellSRefsdal66a}: there are many cosmological
models which expand forever in which, although $\lambda$ and $\Omega$
evolve with time, they never stray very far from their initial values.
In particular, there are many models for which $K$ is roughly $0$ today
and which never deviate very far from this value. Looking at $K$ rather
than just $\Omega$, these models don't suffer from a flatness problem in
the restricted sense, though as mentioned above the same arguments apply
to $\Omega$ (and analogously to $\lambda$) as apply in the $\lambda=0$
case, \ie~they still suffer from the Einstein\ndash de~Sitter problem.
Nevertheless, if, as observations suggest, $K\approx 0$ and
$0<\lnull<1$, then large values of $\lambda$ and $\Omega$ never occur
during the evolution of the universe.  However, this is not really a
satisfactory solution, since it assumes the observed values of \lnull\
and \onull\ rather than explaining why we observe these or similar values
and the Einstein\ndash de~Sitter problem still exists (especially in
respect to $\Omega$ becoming arbitrarily small in the future). 

\begin{figure}
\epsfxsize=\columnwidth
\epsfbox[32 32 496 496]{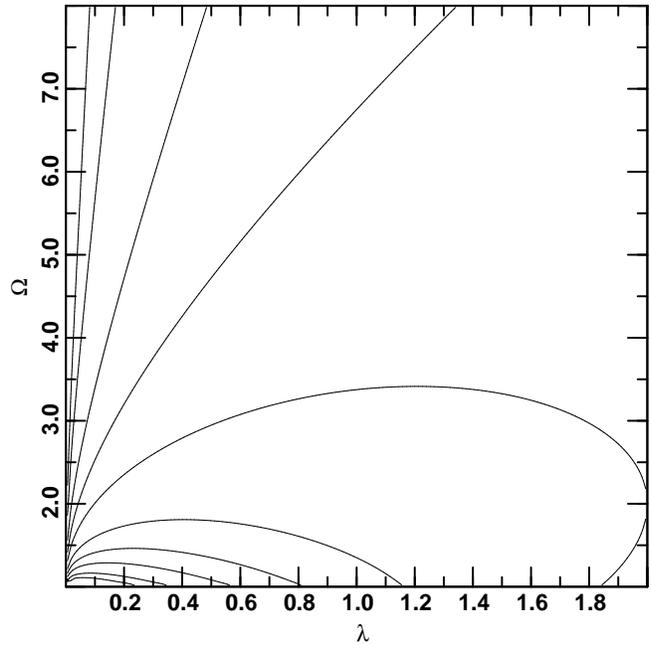}
\caption{The constant of motion $\alpha$ (see \eqn~(\ref{alpha})). From
upper left to lower left, contours are at 0.1, 0.2, 0.5, 1, 2, 5, 10,
20, 50 and 100.}
\label{Lake-fig}
\end{figure}

More interesting is an argument due to \citet{KLake05a} which solves the
Einstein\ndash de~Sitter problem as well for models with $k=+1$ which
will expand forever. (For models which will expand forever, large values
of $\lambda$ and $\Omega$ are possible only for $k=+1$.)  Trajectories
in the $\lambda$-$\Omega$ plane have a constant of motion given by
\eqn~(\ref{alpha}).\footnote{This makes it much easier to plot trajectories
than by calculating $\lambda$ and $\Omega$ as a function of time, though
of course the latter is necessary if one is interested in the amount of
time spent during a particular section of the trajectory.}  It seems
natural to distinguish cosmological models on the basis of their value
of $\alpha$.  Large values of $\lambda$ and $\Omega$ are possible only
for $\alpha \la 1$. This is shown in \Fig~\ref{Lake-fig}.
(Note that, for clarity, only $\Omega>1$ is shown!)  It is obvious that
$\alpha \le 1$ is a necessary condition for having infinitely large
values of $\lambda$ or $\Omega$.  Already for $\alpha=2$ the maximum
value of $\lambda$ is just 2 (for $\Omega=2$) and the maximum value of
$\Omega$ is $\approx 3.5$ (for $\lambda=1.25$). 

In this case, the fine-tuning argument is reversed; only in the case of
fine-tuning do $\lambda$ and $\Omega$ become arbitrarily large.  [If
$\alpha$ is seen as a free parameter which can take on any value between
$-\infty$ and $\infty$, then a `random' value would probably be
arbitrarily large, corresponding to $K \approx 0$ to a very good
approximation.  Interestingly, this is what observations seem to
indicate: $\knull \approx 0$ to within (by now quite good) observational
accuracy, but with no specially interesting values for \lnull\ or
\onull.  For $\knull \approx 0$ and $\lambda>0$, $\lambda$ and
$\Omega$ evolve from $(0,1)$ (the Einstein\ndash de~Sitter model) to
$(1,0)$ (the de~Sitter model, like the Einstein\ndash de~Sitter model a
fixed point, though an attractor rather than a repulsor) without any
large values of $\lambda$ or $\Omega$ along the way; for $\lambda<0$ (in
which case the universe always recollapses), large values do occur but,
as shown above, only during a relatively short time during the history
of the universe.]  This demonstrates quantitatively that there is no
quantitative flatness problem regarding arbitrarily large values of
$\lambda$ or $\Omega$ for models which expand forever.  [Some
models which recollapse but which have $\lambda>0$ also have $\alpha
\approx 1$; this provides an additional argument against the existence
of the flatness problem in these models which complements that made in
\Sect~\ref{collapse} (where, for $\lambda\approx 0$, that argument is
somewhat weaker).]  This argument is also independent of the value of
\hnull.  However, all non-empty models which expand forever
asymptotically approach the de~Sitter model at $(\lambda,\Omega)=(1,0)$.
Thus, one final aspect of the quantitative flatness problem remains:
$\Omega$ can become arbitrarily small.  This is investigated in the next
section. 

If we want to use the tightrope-walker analogy to examine large values
of $\lambda$ and $\Omega$ in models which expand forever, we can think
of the ground representing arbitrarily large values of $\lambda$ and
$\Omega$. However, in this case the tightrope walker is secured by a
safety line of finite length, which prevents him from reaching the
ground (and, as he swings back up after his fall, causes the values of
$\lambda$ and $\Omega$ to decrease).  The only way he can reach the
ground is in the finely tuned case that his rope is long enough, \ie\
infinitely long (since the ground corresponds to arbitrarily large
values of $\lambda$ and $\Omega$). 

Another class of models which expand forever are the bounce models in
which the universe contracts from infinity to a finite size before
expanding forever.  For those near the $\alpha=1$ curve [the A2 curve in
\citet{RStabellSRefsdal66a}], the same arguments apply as for those on
the other side of the curve: only for $\alpha \approx 1$ are large values
of $\lambda$ and $\Omega$ possible for a significant period of time.
These models begin at the de~Sitter model with $\lambda$ and $\Omega$
increasing to $\infty$ in a finite time and return along the same
trajectory in the $\lambda$-$\Omega$ parameter space, in this respect
similar to the models which collapse in the future, thus the argument
against the flatness problem is similar in the two cases.  To
be sure, these models have an infinite extent in time in both
directions, so in this sense there is no flatness problem (or, more
precisely, no `de~Sitter problem', analogous to the Einstein\ndash
de~Sitter problem discussed above) since they are almost always
arbitrarily close to the de~Sitter model.  If we choose starting values
for $\lambda$ and $\Omega$ which are not arbitrarily close to the
de~Sitter model, then the argument is completely analogous to that for
collapsing models.

\section{Cosmological models which expand forever: the weak anthropic 
principle}
\label{anthro}

I have now covered the entire $\lambda$-$\Omega$ parameter space except
for big-bang models with (a) $q<0$ (which implies $\lambda>0$) and (b)
$\Omega$ less than $\approx 2$ (all three values of $k$ are possible)
and shown that in all cases there is no flatness problem.  What about
this remaining portion of parameter space?  Models here all have 
$K\approx 0$ and approach the de~Sitter model asymptotically.  This means
that there is no flatness problem in the restricted sense, as pointed
out by \citet{KLake05a}. However, $\Omega$ becomes arbitrarily small
(and $\lambda$ arbitrarily close to 1).  Thus, there is still a problem
in that we do not observe such values, even though they exist for almost
the entire (infinite) lifetime of the universe. 

This is essentially the question `if the universe lasts forever, then
why are we near the beginning?'  Note that this question could be asked
at any time.  One could leave it at that and say that since any finite
age is arbitrarily close to the beginning, there is nothing special
about our time and thus no flatness problem in the time-scale sense
(\ie~the quantitative flatness problem, why is $\Omega$ not arbitrarily
small today). 

\begin{figure}
\epsfxsize=\columnwidth
\epsfbox[32 32 496 496]{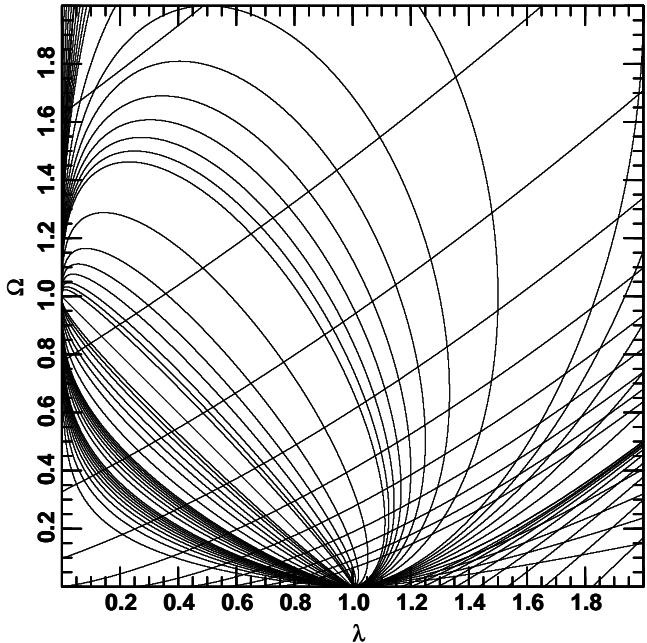}
\caption{Evolutionary trajectories superposed on contours of constant
$Ht$.  Starting at the lower right and moving along a curve
perpendicular to the trajectories, contours for $\alpha$ are at 0.2,
0.4, 0.6, 0.8, 1.0, 1.2, 1.4, 1.6, 1.8, 2, 3, 4, 5, 6, 7, 8, 9, 10, 20,
50, 100, 200, 500, 1000 and 2000.  The straight diagonal line for $K=0$
corresponds to $\alpha=\infty$.  Continuing below this line, the order
is reversed.  From upper left to lower right, contours for $Ht$ are 0.6,
0.7, 0.8, 0.9, 1.0, 1.1, 1.2, 1.3, 1.4, 1.5, $\infty$ (which corresponds
to $\alpha=1$), 1.5, 1.4, 1.3, 1.2, 1.1, 1.0, 0.9, 0.8, 0.7.  Those to
the right of $\infty$ measure the time since the universe was at its
minimum size.} 
\label{anthro-fig}
\end{figure}

One can do better by invoking the weak anthropic principle: cosmological
parameters can be observed only to have values which permit the
existence of cosmologists.  For the first time, the analysis is not
independent of \hnull.  In the de~Sitter model, $H$ is constant in time.
Hence, in models which asymptotically approach the de~Sitter model, $H$
approaches a constant value (which is not that different from \hnull).
If $H$ is similar to the observed value of \hnull, then small values of
$\Omega$ (and values of $\lambda$ near 1) occur only in the relatively
distant future of the universe.  It is difficult to estimate the
probability that cosmologists exist at such future times, but it is
clear that humans in their present form probably won't exist.  In this
sense, we don't observe an arbitrarily small value of $\Omega$ since we
probably wouldn't exist in such a universe.  This is due not to
the small value of $\Omega$ itself but rather to the fact that in the
far future there will be no main-sequence stars \etc.  While this is
only as satisfactory as any use of the weak anthropic principle, (a) this
does not mean that the argument is invalid, (b) it is needed only to rule
out arbitrarily small values of $\Omega$ (and not arbitrarily large
ones, which are ruled out by other arguments above) and, (c) as pointed
out by \citet{KLake05a}, in the restricted sense there is no flatness
problem (\ie~$K \approx 0$ during the entire lifetime of the universe). 
\Fig~\ref{anthro-fig} illustrates this: only arbitrarily old models are
arbitrarily close to the de~Sitter model. 
One can see that arbitrarily small values of $\Omega$ (or values of
$\lambda$ arbitrarily close to 1) occur only when $Ht$ approaches
$\infty$. 

Since the quantity $Ht$ is determined by \lnull\ and \onull,
one could have an arbitrarily young universe arbitrarily close to the
de~Sitter model if \hnull\ were arbitrarily large.  However, such a
universe will spend only a short time near the Einstein\ndash de~Sitter
model, which probably means that structure, and hence cosmologists,
would not form in such a universe. 

[Since $Ht$ has a finite value
(namely $1$) for $\lambda=0$ and $\Omega=0$ and,
for $H$ near the observed value of \hnull, the age of the universe for
$\Omega=0$ would be approximately the same as the observed age,
one might get the impression that, for $\lambda=0$, observing an
arbitrarily small value of \onull\ wouldn't be that unlikely.  However,
this is the case for a universe which is \emph{exactly} described by the
Milne model, which means that it is empty and hence contains no
cosmologists.  The situation is somewhat different when considering the
Milne model as the asymptotic value of a universe with $\lambda=0$ but
$\Omega>0$.  For $\lambda=0$ and $0<\Omega<1$, the universe approaches
the Milne model asymptotically and $H$ approaches $0$. 
In this case, an arbitrarily small $\Omega$ is reached only after an 
arbitrarily long
time, so a similar argument applies as in the case for $\lambda>0$: if
$H$ now is near the observed value of \hnull, then this happens in the far
future when there are no more main-sequence stars \etc.  On the other 
hand, if one assumes
an age near the current observed age but with an arbitrarily small value
of $\Omega$, then this implies that the universe spent only a relatively
short time near the Einstein\ndash de~Sitter model and hence structure
formation would be difficult.]

If we want to use the tightrope-walker analogy to examine the behaviour
at arbitrarily large times in models which expand forever, we can think
of the ground representing the de~Sitter model.  After the tightrope 
walker falls off, he falls towards the ground, but his speed slows with 
time, so that he never actually reaches the ground.

\section{Summary}
\label{summary}

The qualitative flatness problem, \ie~the puzzle why the universe was
arbitrarily close to the Einstein\ndash de~Sitter model\footnote{Or, for
an empty universe, the Milne or de~Sitter model.} at early times, does
not exist.  It is merely a consequence of the way $\lambda$ and $\Omega$
are defined.  Neither does the quantitative flatness problem exist:
although the cosmological parameters in general evolve with time, it is
not puzzling that we don't observe extreme values for them today.  In
the case of models which will collapse in the future this is because
large (absolute) values of $\lambda$ and $\Omega$ occur only during a
relatively short time in the lifetime of such a universe, namely near
the time of maximum expansion.  $\lambda$ and $\Omega$ can become large
only when $H$ becomes small, and this happens only during the time when
the universe is at or near its maximum size.  [Arbitrarily small
(absolute) values, if they occur at all, also occur for only a
relatively short time].  For models which will expand forever, large
values are possible only for $k=+1$.  However, this occurs only for
$\alpha \approx 1$.  In this case, the fine-tuning argument is reversed;
only in the case of fine-tuning do $\lambda$ and $\Omega$ become
arbitrarily large.  Since all models which will expand forever
asymptotically approach $\Omega=0$, arbitrarily small values of $\Omega$
can occur.  Those with $\lambda=0$ (and hence $k=-1$) approach the Milne
model with $\Omega=0$; models with $\lambda>0$, whatever the value of $k$,
approach the de~Sitter model with $\lambda=1$ (the Milne and de~Sitter 
models themselves
are of course stationary points).  (If $\lambda=0$ at any time then
$\lambda=0$ at all times.  Otherwise, arbitrarily small values of
$\lambda$, if they occur at all, occur only for a relatively short
time.)  However, if \hnull\ has a value similar to or smaller than the
observed value, small values of $\Omega$ will occur only in the far
future when anthropic arguments probably make the observation of such a
low value of $\Omega$ unlikely.  While 
(for $\lambda>0$)
a higher value of \hnull\ would
allow a low value of $\Omega$ even for an age near the observed age,
such a universe would have spent only a very short time during which
$\Omega$ was not very small, so structure formation would have been
strongly suppressed. 

It is interesting to note that the three arguments presented here make
it unlikely that we would observe extreme values of \lnull\ or \onull.
This automatically solves the so-called coincidence problem, which has
been called deeply puzzling
\citep*[\eg][]{MTegmarkAVilenkinLPogosian05a}. 

Also interesting is that in the appendix to his seminal paper,
\citet{AGuth81a} anticipates much of the subsequent discussion.  He
points out that even though essentially all cosmological models begin
arbitrarily close to the Einstein\ndash de~Sitter universe (the
qualitative flatness problem), this still leaves the question as to why
the universe is so close to the Einstein\ndash de~Sitter universe today
(the quantitative flatness problem).  He also rejects the idea that some
basic principle must force the universe to conform exactly to the
Einstein\ndash de~Sitter model on the grounds that this is obviously
only an approximation in the case of the real universe (see the
discussion of this above).  He basically recasts the flatness problem as
the longevity problem: fine tuning is required in order that the
universe does not recollapse or thin out to extremely low density within
a very short time.  However, this argument relies on an assumption for
the value of \hnull, while our arguments do not need this assumption
except in the third category.  Also, as I have shown here, even in an
extremely short-lived universe (which of course recollapses), extreme
values of $\lambda$ or $\Omega$ are observed only during a relatively
small fraction of the lifetime of the universe. 

If there is no flatness problem, what does this mean for inflation?  If
there is no flatness problem (or, indeed, even if there were no
motivation at all for inflation), this does not mean that inflation
could not have occurred.  However, it does mean that inflation should
not be taken as a given based on the belief that it explains away the
flatness problem and thus without it classical cosmology leads to absurd
conclusions.  Inflation also solves the monopole and isotropy problems.
However, the monopole problem seems to be more a problem with theories
of particle physics than with cosmology \citep{JNarlikarTPadmanabhan91a}
while \citet{JBarrow95a} claims that there is no isotropy
problem.\footnote{It is, however, debatable whether Barrow's conclusion
is as general as he claims or depends too strongly on his assumptions.} 
If all of these claims are true, then this still does not prove that
inflation didn't happen, but the necessity for inflation or something
like it is weakened if not destroyed altogether.

\section*{\ack}

I thank Nils Bergvall, L\'{e}on Koopmans, Rolf Stabell and Erik
Zackrisson for comments on the manuscript.  Figures were produced with
the \textsc{gral} software package written by Rainer Kayser.

\bibliographystyle{mn2e}
%
%

%

\bsp
\label{lastpage}
\end{document}